\documentclass[12pt,preprint,psfig]{aastex}
\newcommand{\etal}{{\it et al.\ }}

\lefthead{Mart\'\i nez-Delgado \etal}
\righthead{Canis Majoris}

\begin{document}

\title{The closest view of a dwarf galaxy: new evidence on the nature of the Canis Major over-density}

\author{David Mart\'\i nez-Delgado\altaffilmark{1,2}, David J. Butler\altaffilmark{1}, Hans-Walter Rix\altaffilmark{1}, Y. Isabel Franco\altaffilmark{1},  Jorge Pe\~narrubia\altaffilmark{1}, Emilio J. Alfaro\altaffilmark{2} and Dana I. Dinescu\altaffilmark{3}} 
\altaffiltext{1}{Max-Planck Institut fur Astronomie,
K\"onigstuhl, 17 D-69117 Heidelberg, Germany}
\altaffiltext{2}{Instituto de Astrof\'\i sica de Andaluc\'\i a (CSIC), Granada, Spain}
\altaffiltext{3}{Astronomy Department, Yale University, New Haven, USA}

\begin{abstract}
We present a first deep colour-magnitude diagram of the putative  central
region (0.5$\arcdeg \times$ 0.5$\arcdeg$) of the Canis Major stellar over-density $(l,b)=(240,-8)$ found by Martin et al. (2004), which
has been proposed as the remnant of a dwarf  satellite accreted onto
the Milky Way on a near-equatorial orbit. We find a narrow (in apparent magnitude) main-sequence,
extending 6 magnitudes below the turn-off to our limiting magnitude of B$\sim$ 24.5\,mag.
This main sequence has very high constrast ($>$3) 
with respect to the thin/thick disk/halo background; its narrowness at brighter magnitudes clearly 
implies the presence of a distinct and possibly still bound stellar system. We derived the line-of-sight size (r$_{1/2}$) of this system
 based on the B-band width of the lower main sequence, obtaining  0.94 $\pm$ 0.18 (random) $\pm$ 0.18 (systematic)\,kpc. 
That size matches a model prediction for the main body of the parent galaxy of the Monoceros
 tidal stream. The high density contrast and limited spatial extent in the radial direction are very
hard to reconcile with the 
alternative explanation put forward to explain the Canis Major stellar-overdensity:
 a flared or warped Galactic disk viewed in projection (Momamy et al. 2004). We also derived a central surface brightness of $\mu_{V,0}= 23.3 \pm 0.1$\,mag arcsec$^{-2}$  and an absolute magnitude of  $M_{V}=-14.5 \pm 0.1$\,mag.
These values place the Canis Major object in the category of
dwarf galaxy in the the $L_{V}$--size and  $M_V -\mu_{\rm V}$ planes for such
objects.  However, like the Sagittarius dwarf, it is an outlier in  the
$[\rm Fe/H] -M_V$ plane in the sense that it is
too metal rich for its estimated absolute magnitude. This suggests
 that the main mechanism driving its  recent and current star formation
 history  (possibly tidal stripping) is different to that of isolated
 dwarfs.

\end{abstract}


\section{Introduction}

The Milky Way offers a unique laboratory for testing the hierarchical galaxy 
formation scenario through direct evidence of  past  merging and 
tidal disruption events, which result in extensive
stellar streams or large scale substructures in the
Galactic halo or even in the disk (Navarro 2004, and references 
therein). Resolving 
such tidal streams into stars and measuring  the
phase-space coordinates for these stars,  provides a  fossil dynamical accretion record of
unparalleled accuracy. These can be directly compared with N-body simulations
in order to reconstruct their dynamical history and their impact on the Milky
 Way at the present and recent epochs.

Recently, the Sloan Digitized Sky Survey (SDSS) team reported the discovery of a coherent 
giant stellar structure  at low galactic latitudes (Newberg et al. 2002; Yanny et al. 2003), which 
 appears to form a ring around the Milky Way. Since then, there has been a tremendous amount of 
follow-up observational effort probing its structure and kinematics in order to understand its origin (see Majewski 2004 
and references therein). N-body simulations (Martin et al. 2004a;
 Pe\~narrubia et al. 2004) shows that this ring structure can be naturally attributed to  the
 tidal stream of a dwarf galaxy  (named the Monoceros tidal stream, or the Galactic Anticenter 
tidal stream). If this stellar overdensity ring is a tidal tail feature, it must have 
had a ``parent'' galaxy, which may or may not be completely disrupted by now.
The existence and location of such a parent galaxy is still controversial;  
the best candidate is the Canis Major
 (CMa) dwarf galaxy candidate, a strong, ellipsoidal overdensity of red giant stars, 
 discovered in the namesake constellation by Martin et al. (2004a) from an analysis of
the 2MASS survey. Bellazzini et al. (2004) presented a color-magnitude diagram (CMD) 
in the surroundings of the candidate CMa dwarf, concluding that the system is
situated at 8$\pm$ 1 kpc from the Sun and that it is composed of a metal-rich, intermediate-age
 stellar population.  As CMa is only $\sim 8$ degrees from the Galactic plane,
Momany et al. (2004) suggested an alternative interpretation, namely that
this over-density  is  the signature of the Galactic warp in this direction of the sky,
and not a distinct satellite. However, Martin et al. (2004b) reported a narrow
distribution of radial velocities for the center of
this structure,  and argued on this basis for the interpretation of
CMa as an  accreted dwarf satellite remnant, in an orbit near the Galactic plane. 

To help discriminate between these two hypothesis -- accreted dwarf galaxy, or stellar warp -- ,
we present here deep broad-band photometry of the CMa center. As our results confirm 
and greatly strengthen the case that the stellar over-density towards CMa is part of a distinct dwarf galaxy, we refer to it as 'CMa dwarf' 
throughout.

\section{Observations and data reduction}

 Our observations were carried out in $B$ and $R$ 
Johnson-Cousins filters with the 2.2m ESO/MPG telescope at the La Silla Observatory (Chile) in service
 mode during December 14-17th, 2003.  We used the Wide-Field Imager (WFI) installed at the prime focus, which holds eight
 $2046 \times 4098$ pixel 
EEV chips, with a  scale of $0.238''$ pixel$^{-1}$. Our field covers a total area of about $0.25$ deg$^2$ and is centered at galactic
coordinates (l,b)=(240.15, -8.07), which is,
within the uncertainties,  the nominal center of the CMa over-density given by
Martin et al. (2004a). The total
exposure times were 3700s  and 2700s in $B$ and $R$ respectively. A shorter
exposure of 100s in both bands was also taken to recover the brighter part
of the CMD.  

 Bias and flat-field corrections were done with IRAF. DAOPHOT and ALLSTAR (Stetson 1994) were used to
 obtain the photometry of the resolved stars. Aperture corrections were estimated using a set
of about 50 isolated, bright stars in the CMa frame, with a variance of $\sigma\sim 0.001$. Transformation of the instrumental magnitudes
 into the
 standard photometric system  were obtained from observations of  the  Landolt standard field SA95 taking during two  photometric 
nights bracketing our
observations and using the atmospheric extinction coefficients estimated on November 8th, 2003 taken from the WFI homepage. 
 Putting all the errors together, the
total zero-point uncertainty of our photometry can be estimated to be about
$\sigma=0.05$ for $(B-R)$ and $\sigma=0.07$ for individual bands. The resulting catalogue of stars was filtered using the error parameters given
by ALLSTAR, retaining only stars with acceptable CHI and SHARP parameters and $\sigma_{B} <$ 0.2 and $\sigma_{R}< $0.2.
 
Artificial star tests were performed in the usual way (see Aparicio, Carrera
\& Mart\'\i nez-Delgado 2001) using a total of 20000 artificial stars to check the observational effects and estimate the completeness
 factor as a function of magnitude.

\section{The color--magnitude diagram }\label{cmdsec}

\begin{figure}
\epsscale{1.0}
\plotone{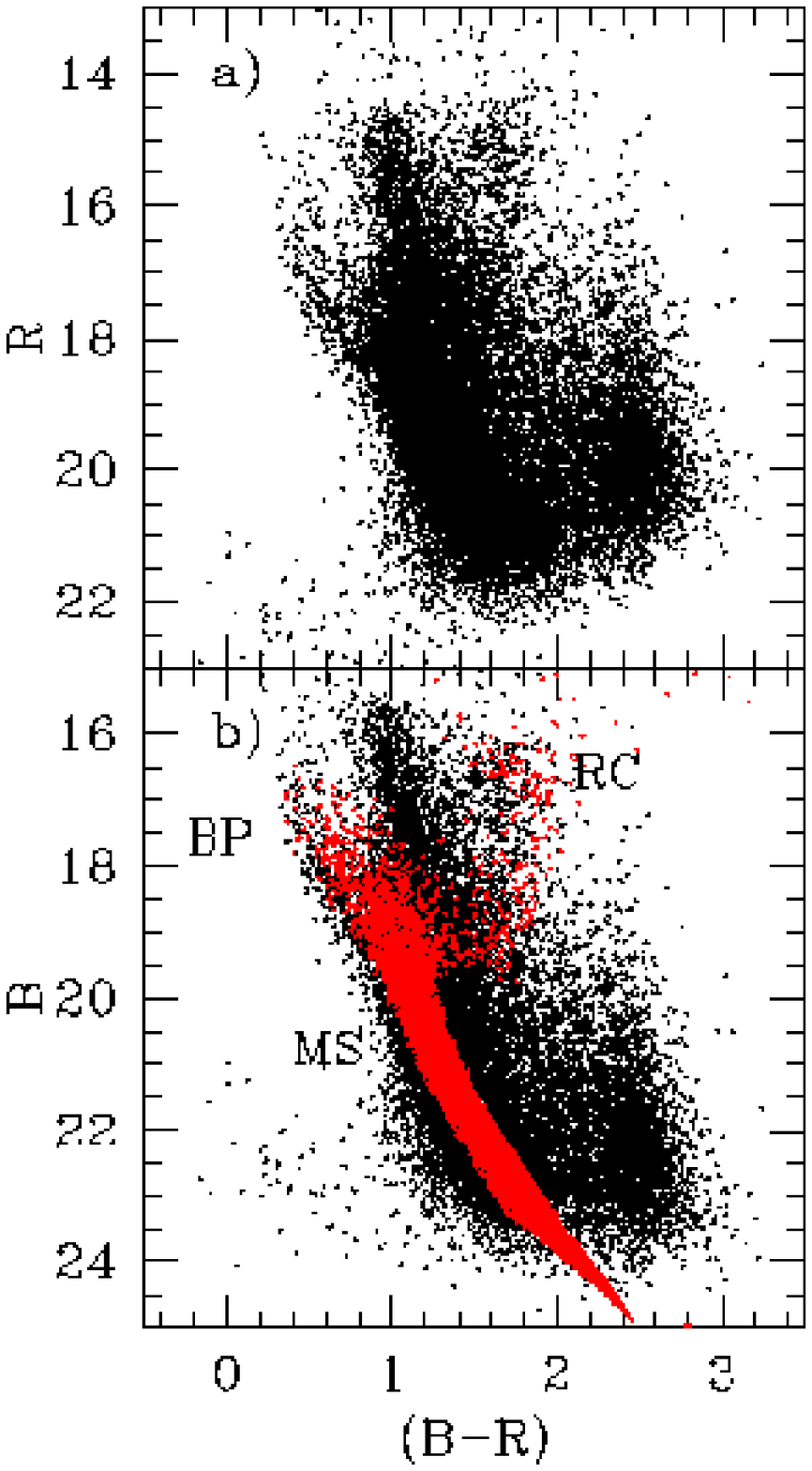}
\caption{(a) $(B-R)$ vs. $R$ colour-magnitude diagrams of the center of the 
 CMa overdensity  based on our short exposure data (see Sec. 3 for discussion). (b)A synthetic CMD (red points) for a metal rich population (see Sec.4.1)  as a possible solution for the CMa  stellar population is overplotted in the observed  $(B-R)$ vs. $B$ CMD (black points). This diagram is only presented here with illustrative purpose in order to provide a better identification  of those regions of the diagram  populated by the brightest stars (red clump (RC), main-sequence (MS) and upper MS (BP) of the putative dwarf galaxy againts the
 Galactic foreground and background contamination. \label{fig1}}
\end{figure}

Fig.~\ref{fig1}a shows the [$B-R$,$R$] CMD of the center of the possible
CMa dwarf based our $t_{exp}=100s$ data.  The stellar densities in this CMD confirm inmediately that the distribution of stars in the Milky
 Way in this
low-latitude direction grossly deviate from the expectations of a smooth 
thin/thick/halo distribution: there is a conspicuous main-sequence (MS) feature
(labelled MS in Fig 1b), with a possible arc-shape turnoff at $R$= 18.48 $\pm$ 0.11 mag . The  $B-R$ colour used on our CMD provides the  separation of the redder plume populated  by thick disk stars (the parallel sequence running from
$[R,(B-R)]\sim$ (15,1.0) to $[R,(B-R)]\sim$ (19,1.6))  from this MS feature, which is about $\sim$ 0.2\,mag bluer
 at $R\sim 18$. Secondly, a second plume of possible MS stars (labelled {\it BP} in Fig. 1b) is observed
 extending brighter and bluer than the MS-turnoff and reaching $R\simeq 16.0$, avoiding the Galactic star 
contamination. This blue extension cannot be reproduced by any star-count models of the Milky Way (see also
Bellazzini et al. 2004) and is similar to those observed in the CMD of
Local Group (LG) dwarf spheroidal (dSph) galaxies (Draco:Aparicio et al. 2001; Ursa Minor: Carrera, Aparicio \& Mart\'\i nez-Delgado 2002). Although it could be produced by blue straggler stars rather than by MS stars of the dwarf galaxy (see discussion in Carrera et al. 2002), the fact that it departs from the MS at $R\sim$ 19.0 (fainter than the MS turnoff) is more compatible with the star formation burst hypothesis than the blue stragglers one, suggesting that the CMa system has had at least two distinct epoch of star formation. The narrow 
distribution of these blue stars strongly suggests that they are all at a
similar distance, which would not expected if the Galactic warp/flare were the origin of this  stellar population.
 Our diagram also shows 
 evidence of  a possible red clump (labelled {\it}RC in Fig. 1b) at $(B-R)\sim$ 1.5 and $R\sim 15$, but a suitable control 
field or radial velocity follow-up is necessary  to confirm this over-density as part  of the CMa dwarf. All the CMD features
described above can be easier identified in Fig~\ref{fig1}b, that shows a synthetic CMD of one of the possible solutions for the stellar population of the
CMa dwarf (see Sec. 4.1) overplotted in our observed [$B-R$,$B$] CMD.\footnote {This synthetic CMD is used here for a qualitative understanding of the expected features in
the CMD of the galaxy population rather than a quantitative comparison with the
data, that is out of the scope of this paper.} The remarkable similarity of the morphology of this  model CMD with the observed one also supports the dwarf nature of CMa.

\begin{figure}
\epsscale{1.0}
\plotone{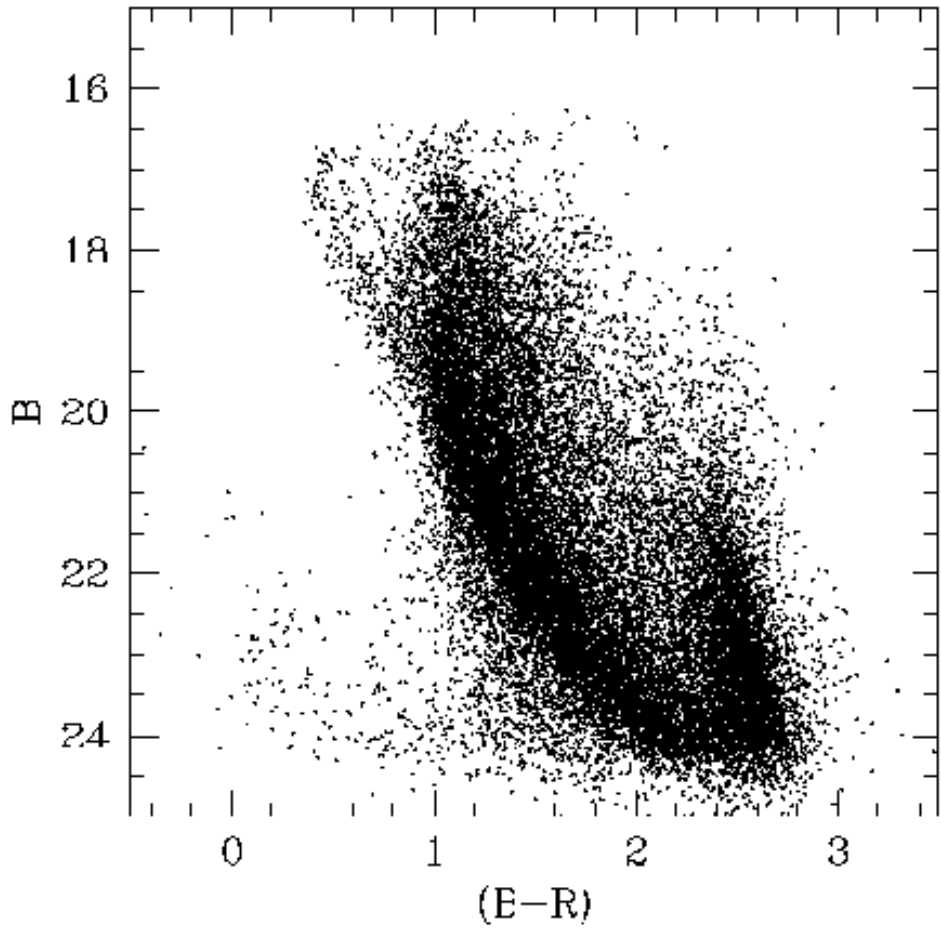}
\caption{  $(B-R)$ vs. $B$ colour-magnitude diagrams of the center of the 
 CMa dwarf galaxy based in our long exposure data. Photometry of stars brighter than $B\sim 20$ mag. suffers larger errors due to
 CCD saturation, but they were included in the diagram as a guide for matching the MS feature in this figure and Fig.~\ref{fig1}b.
  \label{fig2}}
\end{figure}

Fig.~\ref{fig2}  shows the [$B-R$, $B$] CMD based on our long exposure
 data.  This CMD reveals a prominent, high contrast, and well-populated MS 
feature extending beyond our limiting magnitude ($B\sim 24.5$). The colour width of this MS feature remains roughly
 constant along a large magnitude range and is comparable to those observed in the CMD of LG dwarf galaxies
 or massive globular clusters. This evidence strongly confirms the presence of a 
limited range in distance, and  therefore  associated  with a possibly still
 bound stellar system  whose distance line-of-sight size and stellar density
 are investigated in the next section.

\section{Properties of the Canis Majoris dwarf galaxy}\label{CMa_size}

\subsection{Distance}\label{dist}

Our CMD does not  show any unambiguous, convincing signature of the bright, post-MS populations
-- the red clump (RC), horizontal branch, RR Lyrae stars, red giant branch stars -- that must accompany
the clearly detected MS stars. Those post-MS classes of stars have been 
extensively used to constrain the distance and stellar population properties of LG dwarf galaxies.
Consequently, our data are not well suited for  a reliable distance estimate to CMa, 
as there are degeneracies between stellar population assumptions, the reddening and
the distance, when considering only the MS in the CMD. In spite of this degeneracy, we try here to set some limits on the distance to CMa dwarf
 by  fitting the observed MS with synthetic CMDs,  assuming some
different evolutionary scenarios for the galaxy. However, it is important to stress that these are not the only possible solutions and than more data from alternative distance indicators are necessary to confirm the distance reported in
this paper. 

 Adopting  E(B-V) = 0.213$\pm$0.029 mag for our field from the Schlegel, 
Finkbeiner \& Davis(1998; SFD98) dust map, our best model CMD fitting
for an old population (age $>$11\,Gyr, with mean Z=0.006 ) provides an upper limit on the distance modulus of $(m-M)_o=13.6\pm 0.2$ ($d_{\sun}= 5.3 \pm 0.2$ kpc).
For a younger stellar population similar to those reported by Bellazzini et
 al. (2004) (age 4--10\,Gyr, with mean Z=0.006 ), one gets  $(m-M)_o=14.2\pm 0.2$
 ($d_{\sun}= 6.9 \pm 0.3$ kpc). However, it has been suggested that the SFD98 dust maps could overestimate
the actual reddening of our  field (Bonifacio, Monai \& Beers 2000). To check this, we have  obtained E(B-V)  data
for individual stars in the WFI field from  a star-by-star interpolation of the SFD98 dust maps. We find that the E(B-V)  distribution
 is skewed, with about 27\%  of the stars across our CMD are affected by E(B-V) $>$ 0.2. Therefore, we ignore stars assigned 
E(B-V) $>$ 0.3 because they cannot have a significant effect on our distance
  and line-of-sight size estimates. \footnote{Choosing a smaller cut-off threshold does not have a significant effect on the results presented in this paper.}
 For the  mean $\pm$ standard deviation of the remaining
 stars, we obtain E(B-V) = 0.08$\pm$0.07\,mag. This lower reddening value does not permit an acceptable MS fitting for the two former model CMDs, as both produce MS features than are bluer than observed. An acceptable fitting is obtained by increasing the mean metallicity of the galaxy (mean Z=0.01), yielding a  distance modulus of  $(m-M)_o=14.5\pm 0.2$  ($d_{\sun}= 7.9 \pm 0.3$ kpc).
This last value is in very good agreement with previous estimates  
based on different CMD indicators (Bellazzini et al. 2004; Martin et al. 2004b) and it is
adopted as the distance to CMa throughout.

\subsection{The Line-of-Sight ``Depth'' of CMa}\label{CMa_size2}

 To estimate the line-of-sight extent (or ``depth'')  of CMa analyse the observed width 
of the MS, $\sigma_{\rm MS,total,B}$. Near the MS turn-off the MS is steep, and at the faintest
magnitudes measurement errors may dominate the width.  Hence, we consider
the  B-band apparent magnitude distribution 
 of stars in the ($B-R$) color range of 1.50-1.55\,mag, using the 
 long exposure data (see Fig.~\ref{fig2}). We modelled this distribution 
as the linear sum  of two components: 1) the smooth distribution of
 underlying stars due to the Galactic  
warp and thin disk, by a linear function; 2) the CMa MS, by 
a Gaussian function, characterized by a standard deviation $\sigma_{\rm MS,total,B}$. 
Even if all CMa stars were at the same distance,  $\sigma_{\rm MS,total,B}^2$ 
would be non-zero, and could be described by the quadrature sum of three components: the 
 intrinsic MS size due to a range of stellar population
ages and metallicities,\footnote{This was obtained from a synthetic CMD computed for
  a constant and arbitrary star formation rate between 4 and 10\,Gyr and Z =
  0.006 ([Fe/H] =  -0.48\,dex).} plus the contribution to the MS width
 from observational effects ($\sigma_{\rm MS,int, B}$);  
and the differential reddening of the field ($\sigma_{\rm diff.red.,B}$), estimated from the SFD98 dust maps.
We then attribute the remaining width of the MS to the distance distribution
of the CMa stars ($\sigma_{\rm CMa,depth,B}$).
 In brief, we fitted the distribution in Fig. 3
 (in a $\chi^2$ sense)  to a linear function 
at B=20-20.5 and 23.5-24\,mag and 
 fitted a Gaussian function in a 
 non-linear least squares way after subtracting the linear fit.
 The fitting was repeated for two adjacent colour intervals in order to obtain a statistical uncertainty in the MS width. 
Their mean and standard deviation data are given in Table~\ref{tbl-1}, together with the rest of the budget used to determine the 
line-of-sight size of CMa.  

Our best fitting $\sigma_{MS,total,B}$, converted into kilo-parsecs, yields $\sigma_{\rm CMa, depth,B}$=0.8$\pm$0.15\,kpc
 or  FWHM = 1.95\,kpc (see Table~\ref{tbl-1}). Based on 2MASS data, Martin et al. (2004b) report that angular extent of the
 over-density corresponds to a
FWHM of about $\sim$ 4.2\,kpc (at a helio-centric distance of
7.1$\pm$1.3\,kpc), significantly larger than the
  line-of-sight value estimated here. This is not necessarily 
 inconsistent, because severe tidal perturbations may cause a tangential streching of the CMa dwarf. In addition, the estimated size of CMa dwarf can provide a reliable estimation of its mass under the assumption that bound stellar systems undergoing disruption have a spatial extension similar or larger than the Jacobi limit, as Pe\~narrubia et al (2004) find for the Monoceros stream progenitor. In that case (Binney \& Tremaine 1986; equation 7-84), $$M_s \sim 3 M(<R_{gal}) (R_{t} / R_{gal})^3 $$
where $M(<R_{gal})$ is the Galaxy mass within $R_{gal}$ and $R_t$ is the
Jacobi limit (also called tidal radius). Adopting $R_t\simeq 1.6$ (which
accounts approximately for 96\% of the dwarf mass) and
the mass profile used by Pe\~narrubia et al. (2004) we find
$M_s> 5\times 10^8 M_\odot$, indicating that the present mass of the CMa
dwarf should be similar to those predicted by our model of the Monoceros
stream progenitor ($3\times 10^8 M_\odot$: Pe\~narrubia et al. 2004).

Finally, in order to compare the size of CMa with those of known dwarf galaxies, we estimate the
line-of-sight half-brightness radius, r$_{\rm 1/2}$, from the Gaussian model of the line-of-sight profile given
in Fig.3.  We obtain   r$_{\rm 1/2}$=0.94 $\pm$ 0.18 (random) $\pm$ 0.18
(systematic)\,kpc. This size is in agreement with the model prediction (Pe\~narrubia et al. 2004); but is 
significantly bigger than that of several dwarf galaxies in the LG (Irwin \& Hatzidimitriou 1995; their Table 4). 

\subsection {Surface brightness and absolute magnitude}

We estimate the central surface brightness (SB)  of the
CMa dwarf by matching the observed surface density of MS stars to scale 
those derived from a (candidate) synthetic CMD for the CMa galaxy (2.5-15\,Gyr; Z=0.006)  for which the
integrated magnitude is known. We used the short exposure data
 and counted stars on the upper MS in a small box such that 
   1$< B-R  <$ 1.5 and 18 $< R <$19;
   the position of this CMD box allows a significant sample of the dominant CMa stellar population. To correct
the MS star-counts for contamination of Milky Way stars, we use the CMD of a control field obtained at
the galactic position (l,b)=(240,+15), taken from our recent survey to investigate the extent and
stellar population of the CMa dwarf (Butler et al., in preparation). We used star counts from the red plume
 region of thick disk MS stars (where 0.5 $<$ B-R $<$ 1.5 and  15 $<$ R $<$ 16.5)  in the control and CMa fields 
 to subtract a scaled contamination level from the CMa MS star
selection box mentioned above. We find $\mu_{\rm V, 0}$ =   23.3
$\pm$ 0.1\,mag\,arcsec$^{-2}$ (taken as the intensity-weighted
average of $\mu_{\rm B, 0}$ and  $\mu_{\rm R, 0}$), 
where the SB uncertainty  was estimated by taking Poisson statistical errors in the star counts. That
 value is very similar to those of Milky dSph satellites (Mateo 1998), and provides a further evidence 
 that we have detected a part of the main body of a  dwarf galaxy and not a piece of a possible 
tidal stream, whose typical SB would  be expected to lie in the range of 30.0 to
 31.5  mag\,arcsec$^{-2}$ (Johnston et al. 1999).

To  estimate the total V-band luminosity, we use an exponential SB
 profile with a (near-IR)
scale length of 0.73$\pm$0.05\,kpc (Martin et al. 2004a) with the mean 
  heliocentric distance taken to be 7.1\,kpc. We obtained
$M_{V}$= -14.5$\pm$0.1\,mag,  corresponding to a total V-band intensity log
 (I$_{V}/I_{V, \odot}$) =7.7$\pm$0.1. This yields a total mass for the satellite of $M/M_{\odot}$ =5
  $\times$  10$^{7}$$\Gamma$ (where $\Gamma=(M/M_{\odot})/(L/L_{\odot}$)). Assuming the mass-to-light ratio of CMa is in the range  $4  < \Gamma < 22 $,
 in solar units, (which contains the majority of
the LG dSphs), the present remnant of the CMa dwarf would have a total mass of
 $2.0\times 10^8 <M/M_{\odot} < 1.1 \times 10^9$, consistent with the value obtained from  the observational 
extension of the satellite and from  the prediction of our theoretical
 simulations (see Sec.~\ref{CMa_size2}). The combination of this luminosity with the size
 derived in Sec.~\ref{CMa_size}
  places  this stellar system among the dwarf galaxies
 in the the $L_{V}$--size plane (see Pasquali et al. 2004; their Fig. 5).
  In addition, CMa follows the well-defined 
 $M_V -\mu_{\rm V}$  relationship shown by Caldwell et al. (1999) for dwarf galaxies. However, assuming the stellar component
of the galaxy has an average metallicity of $[Fe/H]\sim -0.4$ (Bellazzini et al. 2004), this galaxy is,
 like the Sgr dSph, an outlier of  the
 well-known $[Fe/H] -M_V$ relation, since it is too metal rich for its estimated absolute magnitude. This suggests, as in the
case of Sgr, that a different mechanism (possibly tidal stripping) is predominantly driving its star formation history.


 Derived parameters for the CMa overdensity are given in Table~\ref{tbl-2}.

\section{CONCLUSIONS}

From our data we have established that the putative center of CMa is a very high 
contrast density feature that as has very narrow radial extent, $r_{1/2}/R_{G.C.} < 0.1$.  
These results strongly support the interpretation of a distinct, possibly still bound stellar system whose  properties 
are consistent with  those expected for the remnant of a partially disrupted dwarf satellite. 
In turn, the high density contrast and the limited line-of-sight extent of
the CMa would be very difficult to be reconciled (if at all) with a
flared or warped  Galactic disk viewed in projection (Momamy et al. 2004). Although the obtained properties for this satellite (Table 2) in this study 
are subject to considerable uncertainties (stellar population, Galactic contamination, radial profile), they
 are in agreement with those of the known Milky Way dSph satellites (Mateo 1998) and with the known size-luminosity relation followed by Milky Way dwarf spheroidals. 

At a distance of 8 kpc, CMa is the closest dwarf galaxy known. It may not only be the parent of the Galactic low-latitude stellar stream, but also a unique laboratory for
testing galaxy evolution theories. In particular, our CMD shows the
availability  for first time of thousands of MS
 stars of different ages in a magnitude range suited to high resolution
 spectroscopy-based abundance studies using  4-meter class telescopes. This
 will allow one to study the SFH of a dwarf galaxy with
 unprecedented  spectral 
resolution, and is therefore a top candidate for chemo-dynamical
 modelling of dwarf galaxy evolution in the Galaxy.

\acknowledgements

We thank A. Robin, E. D. Skillman and the referee H. Rocha-Pinto for their useful comments. DMD devotes this work 
to the  memory of his grandfather Manuel Delgado-Membrives.

\newpage

\begin{figure}
\epsscale{0.9}
\plotone{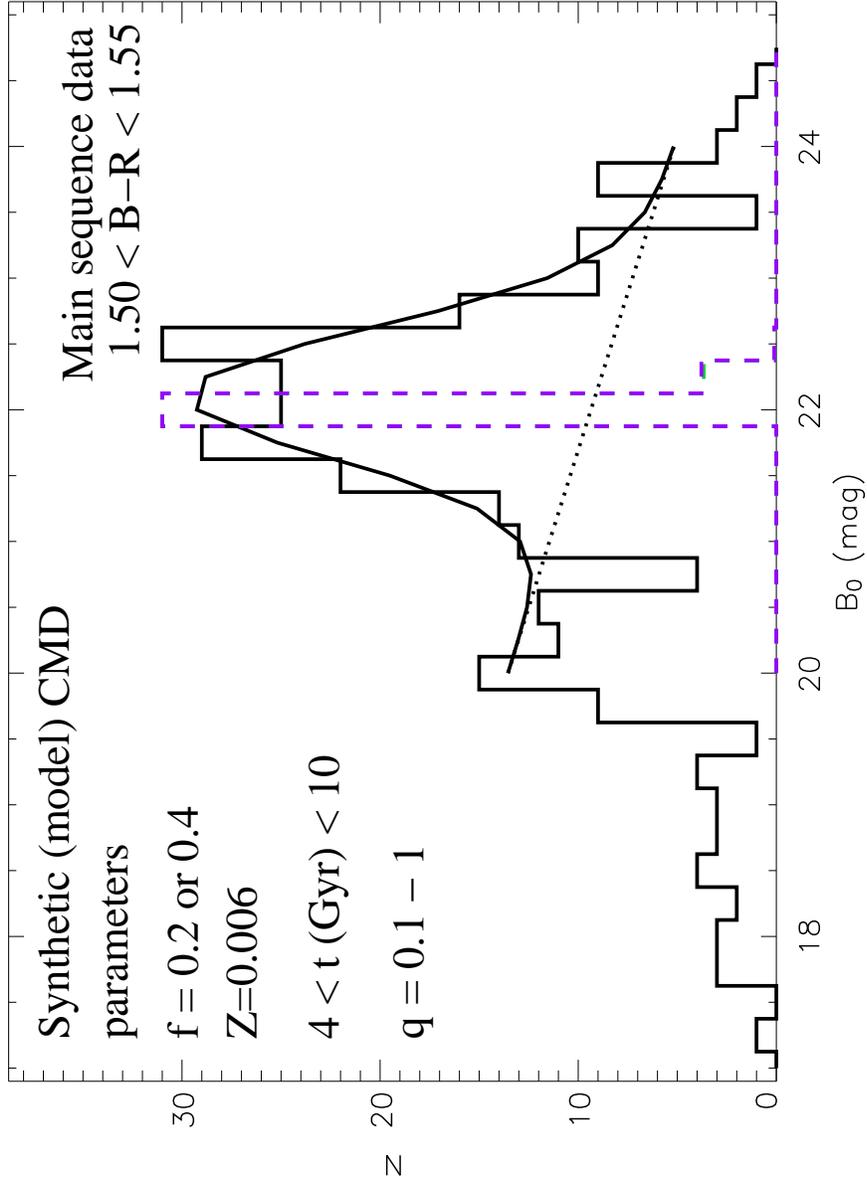}
\caption{B-band histogram derived for the $B-R$ range  1.50-1.55\,mag in
the long exposure data. The corresponding histogram from a
 synthetic CMD (4-10\,Gyr; Z = 0.006), is overlaid (dashed).
This synthetic CMD was computed by means of the IAC-STAR interface (Aparicio \& Gallart
 2004) using  the evolutionary library from Girardi et al. (2000);
 and considered a  binary fraction (f)  of 0.2
 and 0.4  with a flat  probability distribution of
mass ratios (q) between  0.1 and 1.
  The model  distributions are indistinguishable.
The overplotted bell-shaped curve (solid line) is the linear sum of a linear fit to the
 underlying Milky Way stars (dotted line) and a Gaussian fit to the
 remaining (i.e. CMa MS) histogram. See the text
 in Section~\ref{CMa_size2} for
 further details on data fitting and the MS width. \label{fig3}}
\end{figure}

\clearpage

\begin{table}
\begin{center}
 \caption{Budget used to determine the thickness of the CMa MS and an MS model
   with Z=0.006 and  age of 4 to 10\,Gyr\label{tbl-1}}
\begin{tabular}{lr}
\tableline\tableline
Total observed main sequence width \\
 $\sigma$$_{\rm MS,total,B}$ (mag) & 0.57 $\pm$ 0.06 \\
\tableline
Random errors  \\
$\sigma_{\rm MS,int,B}$ (mag)   & 0.19$^a$ \\
 $\sigma_{\rm diff. red., B}$ (mag) & 0.29$^b$ \\
\tableline
Systematic error\\
model MSTO (adopted error) & 0.1 \\ 
Photometry calibration &  0.07 \\
\tableline
\end{tabular}
\tablenotetext{a}{Observational effect estimated from artificial star tests, based on a single age population,
 and determined at B$\sim$22.1\,mag and B-R = 1.5 to 1.55\,mag.}
\tablenotetext{b}{Adopted the Cardelli, Mathis \& Clayton. (1989) reddening law and E(B-V)=0.08\,mag (see Sec.4.1)}
\end{center}
\end{table}

\begin{table}
\begin{center}
 \caption{Relevant properties of the CMa field. \label{tbl-2}}
\begin{tabular}{lc}
\tableline\tableline
 l   (deg) & 240.15 \\
 b  (deg)  & -08.07 \\
 E(B-V)   (mag) & 0.08$\pm$0.07  \\
 d (kpc) & 7.9 $\pm$ 0.3$^a$ \\
\tableline
 $\Delta $d  (kpc) & 1.60$\pm$ 0.30$^b$\\
 r$_{1/2}$ (kpc) &   0.9$\pm$ 0.3 \\
\tableline
 $\mu_{V, 0}$ (mag\,\,arcsec$^{-2}$)  &  23.3$ \pm$ 0.1$^c$ \\
\tableline
\end{tabular}

\tablenotetext{a}{Model used: Z=0.001, 4-10\,Gyr, f=0.2, q=0.1-1}
\tablenotetext{b}{2$\times$$\sigma$ line-of-sight size}
\tablenotetext{c}{Model used: Z=0.006, 4-10\,Gyr, f=0.2, q=0.1-1}
\end{center}
\end{table}


\begin{thebibliography}{}

\bibitem[Aparicio]{Aparicio} Aparicio, A., Gallart, C. 2004, \aj, 128, 1465

\bibitem[Aparicio]{Aparicio} Aparicio, A., Carrera, R., Mart\'\i nez-Delgado, D. 2001, \aj, 122, 2524

\bibitem[Bellazzini]{Bellazzini} Bellazzini, M., Ibata, R., Monaco, L., Martin, N., Irwin, M. J., Lewis, G. F. 2004, MNRAS, 354,1263

\bibitem[Binney]{Binney} Binney, J, Tremaine, S., Galactic Dynamics,Princeton
University Press, Princeton, New Jersey

\bibitem[Bonifacio(2004)]{bonifacio04} Bonifacio, P., Monai, S., Beers, T. C.,  2000, AJ, 120,2065

\bibitem[Caldwell(1999)]{caldwell, N.} Caldwel, 1999, AJ, 118, 1230


\bibitem[Cardelli(1989)]{cardelli89}Cardelli, J. A., Clayton, G. C.,
    Mathis, J. S., 1989, ApJ, 345, 245
\bibitem[Carrera]{carrera} Carrera, R., Aparicio, A., Mart\'\i nez-Delgado, D., Alonso-Garc\'\i a, J. 2002, AJ, 123, 3199
\bibitem[girardi]{girardi} Girardi, L., Bressan, A., Bertelli, G., Chiosi, C. 2000, A\&AS, 141,371

\bibitem[irwin]{irwin} Irwin, M. \& Hatzidimitriou, D. 1995, MNRAS, 317, 831

\bibitem[johnston]{irwin} Johnston, K. V., Zhao, H., Spergel, D. N., Hernquist, L. 1999,\apj, 512, 109

\bibitem[Navarro]{navarro} Navarro, J. F. 2004, in {\it Penetrating Bars through
Marks of Cosmic Dust'}, in press (astro-ph/0405947) 
\bibitem[Majewski]{majewski} Majewski S. R. 2004, in {\it Satellites and Tidal Streams}, Prada, Mart\'\i nez-Delgado \&
Mahoney eds., San Francisco, ASP, ASP. Conference Series vol. 327, 63

\bibitem[Martin(2004)]{martin04}Martin, N. F., Ibata, R. A.,  Bellazzini, M., 
 Irwin, M. J.,  Lewis, G. F.,  Dehnen, W., 2004a, MNRAS, 348, 12 
 
\bibitem[Martin(2004)]{martin04} Martin, N. F., Ibata, R. A.,  Conn,
  B. C.,  Lewis, G. F.,  Bellazzini, M.,  Irwin, M. J.,  McConnachie, W., 2004b, MNRAS, 355, 33


\bibitem[Mateo(1998)]{mateo98}  Mateo, M. L., 1998, ARAA, 36, 435
\bibitem[Momany]{momany}  Momany, Y., Zaggia, S. R., Bonifacio, P., Piotto, G., De Anfeli, F., Bedin, L. R., 
Carraro, G. 2004, A\&A, 421, L29
\bibitem[Newberg]{newberg} Newberg, H. J. et al. 2002, ApJ, 596, L191


\bibitem[pasqualy]{pasqualy} Pasquali, A., Larsen, S., Ferreras, I., Gnedin, O. Y., Malhotra, S., Rhoads, J. E., Pirzkal, N., Walsh, J. R. 2005, AJ, 129, 148
\bibitem[Penarrubia et al. 2004]{pietr04} Pe\~narrubia, J., Mart\'\i nez-Delgado, D., Rix, H.-W., G\'omez-Flechoso, M. A., Munn, J., Newberg, H. J., Bell, E. F., Yanny, B., Zucker, D., Grebel, E. K. 2004, ApJ, in press (astro-ph/0410448)

\bibitem[SFD98]{sfd98} Schlegel D. J., Finkbeiner, D. P., Davis, M., 1998, ApJ,
  500, 525 (SFD98)
\bibitem[Stetson]{stetson} Stetson, P. B. 1994, PASP,106, 205
\bibitem[Yanny et al. 2003]{Yanny} Yanny, B. et al. 2003, ApJ, 588, 824

\end{thebibliography}
\end{document}